\documentclass[manuscript]{aastex}

\shorttitle{Crystallization condition of amorphous silicate}
\shortauthors{Tanaka, Yamamoto, and Kimura}

\begin{document}

\newcommand{\sub}[1]{_{\mbox{\scriptsize {#1}}}}
\def\simg{\hspace{1ex} ^{>}\hspace{-2.5mm}_{\sim} \hspace{1ex}}
\def\siml{\hspace{1ex} ^{<} \hspace{-2.5mm}_{\sim} \hspace{1ex}}


\title{ Low-Temperature Crystallization  
 of   Amorphous Silicate \\ in  Astrophysical Environments 
 }

\author{Kyoko. K. Tanaka\altaffilmark{1}, Tetsuo Yamamoto\altaffilmark{2},}

\affil{Institute of Low Temperature Science, Hokkaido University, 
    Sapporo 060-0819, Japan}


\and

\author{Hiroshi Kimura\altaffilmark{3}}
\affil{Center for Planetary Science, Kobe 657-8501, Japan} 


\altaffiltext{1}{Research Fellow, 
Institute of Low Temperature Science, Hokkaido University}
\altaffiltext{2}{Professor,
Institute of Low Temperature Science, Hokkaido University}
\altaffiltext{3}{Associate Professor, Center of Planetary Science,
 Kobe University}


\begin{abstract}

  We construct a theoretical model for low-temperature crystallization
  of amorphous silicate grains induced by exothermic chemical reactions.
  As a first step, the model is applied to the annealing experiments, in
  which the samples are (1)~amorphous silicate grains and (2)~amorphous
  silicate grains covered with an amorphous carbon layer.  We derive the
  activation energies of crystallization for amorphous silicate and
  amorphous carbon from the analysis of the experiments. Furthermore, we
  apply the model to the experiment of low-temperature crystallization
  of amorphous silicate core covered with an amorphous carbon layer
  containing reactive molecules.  We clarify the conditions of
  low-temperature crystallization due to exothermic chemical reactions.
  Next, we formulate the crystallization conditions so as to be
  applicable to astrophysical environments.  We show that the present
  crystallization mechanism is characterized by two quantities: the
  stored energy density $Q$ in a grain and the duration of the chemical
  reactions $\tau$.  The crystallization conditions are given by $Q >
  Q\sub{min}$ and $\tau < \tau\sub{cool}$ regardless of details of the
  reactions and grain structure, where $\tau\sub{cool}$ is the cooling
  timescale of the grains heated by exothermic reactions, and
  $Q\sub{min}$ is minimum stored energy density determined by
  the activation energy of crystallization.  Our results suggest that
  silicate crystallization occurs in wider astrophysical conditions than
  hitherto considered.

\end{abstract}





\section{Introduction}

Infrared observations of various comets suggest the existence of
crystalline silicates in their nuclei \citep{bregman1987,
  molster1999}.  In contrast, silicate in the interstellar medium is
 almost or entirely amorphous, namely, free from crystallites
\citep{li2007}.  It is, therefore, popular to assume that crystalline
silicates observed in comets formed in the inner solar nebula  (e.g.,
Gail 2001; Bockelee-Morvan et al. 2002; Harker and Desch 2002;
 Keller and Gail 2004).
This is out of harmony with the composition of the gas in cometary
comae, observations of which indicate the preservation of interstellar
ices in the cold outer nebula \citep{biermann1982, mumma1996}.  As
noticed by \citet{huebner2002}, none of the thermal mechanisms
proposed for silicate crystallization allows comets to retain the
interstellar composition of ices in their nuclei.

Besides comets, infrared spectra tell us the presence of crystalline
silicates in various kinds of object such as AGB stars, post-AGB
stars, red supergiants, Herbig Ae/Be stars, and protoplanetary disks
around young stellar objects. It is known that crystallization of
amorphous silicate due to annealing requires temperature $T$ above
1000\,K \citep{hallenbeck1998, fabian2000, murata2007}.  It is
generally considered that silicate does not crystallize below its
glass transition temperature, which is 990 K for forsterite
composition \citep{speck2008}.  However, infrared spectra of dust
shells around evolved oxygen-rich stars exhibit the presence of
several emission features of crystalline silicates in their cool dust
shells ($T < 300\,$K), while silicates condense in dusty outflows with
amorphous structure as evidenced by the spectra \citep{waters1996}.
Crystalline silicates are also observed in ULIRGs (ultraluminous
infrared galaxies), implying that they are located in the cool, outer
regions \citep{spoon2006}.  From these observational results, several
researchers claim a necessity of a yet unknown crystallization process
in low-temperature environments \citep{waters1996, molster1999,
  molster2001, spoon2006, watson2009}.

Physically, crystallization is a re-arrangement of atoms irregularly
placed in a solid so that the atoms occupy lattice sites; in such a
configuration the solid attains the lowest internal energy. However,
the atoms must overcome the energy barrier, namely, activation energy
of crystallization  $E$. In annealing, a fluctuation of thermal energy is
used to overcome the energy barrier. However, a fluctuation of the
thermal energy is not necessarily a requisite for crystallization.
\citet{carrez2002} and \citet{kimuray08} showed that amorphous
silicate with forsterite composition crystallizes at room temperature
when they are irradiated by electrons in a transmission electron
microscope (TEM).

Recently, we have proposed another mechanism of low-temperature
crystallization due to exothermic chemical reactions of reactive
molecules in an organic refractory mantles surrounding an amorphous
silicate core \citep{yamamotochigai, yamamoto2009a, yamamoto2009b}.
Once the reactions are triggered by moderate heating above a few
hundred kelvins, chain reactions of reactive molecules contained in
the organic mantle  release heat that crystallizes the surface
layer of the amorphous silicate core. We estimated the degree of
crystallinity to be 0.4 to 20\,\% in volume for the concentration of
reactive molecules of 1 to 10\,\%.   It was shown that the degree
  of crystallinity was sufficient to reproduce the observed strength
  of infrared features characteristic of crystalline silicates in
  cometary comae \citep{kimura08, kimura2009}. 

Nevertheless, we notice that the previous model of
\citet{yamamotochigai} and \citet{yamamoto2009b} is incomplete, 
  because the model permits crystallization at temperatures above the
  melting point, although, in fact, the existence of crystalline
  silicate is not allowed in these temperatures.  If one excludes
crystallization at temperatures higher than the melting point, one
would have a lower volume fraction of crystalline silicate.  
  Furthermore, their model neglects the effect of finite particle
  size, namely, neglects accumulation of the heat of chemical
  reactions in the particle of a finite size. In consequence, we
  expect that the particle-size effect increases the degree of
  crystallinity.  These two points should be addressed in the model
to properly examine the plausibility of nonthermal crystallization due
to exothermic chemical reactions.

In this study, we construct a model for low-temperature
crystallization of an amorphous silicate core coated  with a
layer of carbonaceous material, by taking into account the melting
point and  a finite particle  size  (in section 2).   In
  view of the analyses of cometary dust and primitive interplanetary
  dust particles, we consider a grain having an amorphous silicate
  core and a mantle of carbonaceous material.  Amorphous carbon is one
  of the main components in the carbonaceous material. 
  \citep{keller1996, keller2000, kissel1987, jessberger1999}.  We
determine the activation energies of crystallization in amorphous
silicate and amorphous carbon using the revised model in comparison
with two previous crystallization experiments, in which particles of
(1)~amorphous silicate and (2)~amorphous silicate covered with a
carbonaceous layer are annealed.  These evaluations enable us to
derive quantitative crystallization conditions. We demonstrate the
validity of the model by its application to crystallization
experiments due to exothermic chemical reactions demonstrated by
\citet{kaito2007a} in section 3.   Finally we  formulate the
crystallization conditions applicable to various astrophysical
environments in section 4.  We discuss the feasibility of nonthermal
crystallization in astrophysical environments and summarize our
conclusions in section 5.

\section{Model}

\subsection{Basic equations} 


Cometary dust and interstellar dust are well modeled as aggregates of
small grains having an amorphous silicate core and an organic refractory
mantle \citep{kimura03a}. In-situ measurements of mass spectra for dust
in a coma of comet 1P/Halley show evidence of such a core-mantle
structure, in which a core and a mantle are composed of silicate and
organic refractory material, respectively \citep{kissel1987}. A cluster
of submicron grains with amorphous silicate enclosed within carbonaceous
material such as amorphous carbon rich in organic material is common in
primitive interplanetary dust particles of cometary origin
\citep{keller1996, keller2000, kissel1987, jessberger1999}.  
Because silicate crystallization in each constituent grain of 
the cluster occurs independently, we hereafter consider a single particle
having a silicate core covered by amorphous carbonaceous material.
 We expect that reactive molecules have been formed by the exposure
of ultraviolet radiations and high energy particles in molecular clouds
and stored in the organic layer of interstellar dust.  

Once the dust is heated up, the diffusion rate of reactive molecules
increases and chemical reactions are triggered in the mantle layer. The
energy released by the reactions raises the temperature and in turn
expedites further reactions.  As a consequence of the heat flow, the
amorphous silicate core crystallizes.  If the mantle layer is
composed of amorphous carbon, the amorphous carbon may also be
graphitized (Kaito et al. 2006,  see also Speck et al. 2009).  
 We take into account two types of crystallization of the silicate
core. 

 The timescale of heat transfer in a particle is estimated to be
\begin{eqnarray}
\tau\sub{heat} 
= r^2/\chi 
= 3 \times 10^{-8} 
    \left( \chi \over 10^{-3} \,{\rm cm^2\, s^{-1}} \right)^{-1}
               \,{\rm s}
\end{eqnarray}
 for the particle radius of $r = 50\,$nm.  The values of $\chi$
   range from 4 to $20 \times 10^{-3}\,{\rm cm^2\, s^{-1}}$ depending
   on the temperature, compositions, and degree of crystallization
   \citep{whittington2009,hofmeister2009}.  Thus we find
   $\tau\sub{heat} < 10^{-8}\,$s, which is much shorter than the
 crystallization timescale $\simg 10^{-5}$\,s (see Appendix A).
 Consequently, it is safe to assume that a particle is isothermal
 during the crystallization process.

The energy budget of the particle at temperature $T$ is described by
\begin{eqnarray}
{4 \over 3} \pi (a+h)^3 \rho\sub{d} c\sub{p} {d T \over dt}
 = {4 \over 3} \pi [(a+h)^3-a^3] \dot\varepsilon
 + \Gamma\sub{si} + \Gamma\sub{c}
 - \Lambda\sub{rad}-\Lambda\sub{coll}, 
\label{exp3-eq}
\end{eqnarray}
where $a$ is the radius of the silicate core, $h$ is the thickness of
the carbonaceous mantle, $\rho\sub{d} (\simeq 3.0$\,g\,cm$^{-3}$) is
the mean density of the particle, and $c\sub{p} (= 1.5 \times
10^{7}\,$erg\,g$^{-1}$\,K$^{-1}$) is the mean specific heat of
amorphous carbon and silicate \citep{navrotsky1995, lide1996}.  The
first term on the right-hand side in Eq.~(\ref{exp3-eq}) is a heating
term due to exothermic chemical reactions and $\Gamma\sub{si}$ is a
heating rate due to deposition of the latent heat of crystallization
of amorphous silicate, and $\Gamma\sub{c}$ is that of amorphous
carbon.  For cooling processes, we take into account two cooling
mechanisms due to thermal radiation $\Lambda\sub{rad}$ and collisions
of gas molecules surrounding the particle, $\Lambda\sub{coll}$. We
assume a first order reaction, namely, the main reactions are those of
reactive molecules contained in the mantle with the mantle materials.

Then the time variation of the number density $n\sub{A}$ of the reactive
molecules is expressed by
\begin{eqnarray}
 { d n\sub{A} \over dt} = -\frac{n\sub{A}}{\tau},
\end{eqnarray}
where $\tau$ is a timescale of the reactions. 
The heating rate $\dot\varepsilon$ by the reactions is given by
\begin{eqnarray}
 \dot\varepsilon
 = - { d n\sub{A} \over dt} q\sub{r}
 = {k Q \over \tau} \exp\left(-{t \over \tau} \right).
\end{eqnarray}
Here, $Q $ is the energy density stored in the carbonaceous layer given
by $Q = n\sub{A,0} q\sub{r}/k$, where $n\sub{A,0}$ is the initial number
density of reactive molecules, and $q\sub{r}$ is energy release per
reaction.
  The cooling rate due to
thermal radiation, $\Lambda\sub{rad}$, is given by
\begin{eqnarray}
 \Lambda\sub{rad} = 
  4 \pi (a+h)^2 \sigma\sub{B} \epsilon (T^4-T_0^4) ,
\end{eqnarray}
where $\epsilon$ is the efficiency of thermal emission from the surface
of the carbonaceous mantle, $T_0$ is the ambient radiation temperature,
and $\sigma\sub{B}$ is the Stefan-Boltzmann constant.  The cooling rate
$\Lambda\sub{coll}$ due to collisions of ambient gas molecules of
temperature $T_0$ is given by
\begin{eqnarray}
\Lambda\sub{coll}
 = 2 \pi (a+h)^2  n \bar v k (T- T_0).
\end{eqnarray}  
Here, $n$ is the number density of the gas molecules, $k$
  is the Boltzmann constant, and $\bar v =
\sqrt{8 k T_0/\mu m_{\rm a}}$ is their mean thermal velocity,
where $\mu $ is the mean molecular weight and $m_{\rm a} = 1.6 \times
10^{-24}\,{\rm g}$ is atomic mass unit. 

We assume that the crystallization proceeds from the interface between
the core and the mantle as is supported by crystallization experiments
of grains having a silicate core and an amorphous carbon mantle (see
section 2.2).  This means that a crystal growth occurs through
heterogeneous nucleation on the interface.  Namely, crystallization of
an amorphous silicate core proceeds inward in the core from the
interface and graphitization of an amorphous carbon mantle proceeds
outward in the mantle from the interface.

Denoting the distance of the silicate crystallization front by $a_{\rm
si}$ and the graphitization front from the center of the particle by
$a_{\rm c}$, equations of the crystal growths in the silicate core and
the carbonaceous mantle are given, respectively, by
\begin{eqnarray}
 {d a\sub{si}\over dt} &=&
 -\Omega\sub{si}^{1/3} \nu\sub{si}
 \exp\left(-{E\sub{si} \over k T  } \right)
 \left[1- \exp\left(-{q\sub{l,si} \Delta T\sub{si}
  \over kT^2} \right) \right],  
\label{growth-si}\\
 {d a\sub{c}\over dt} &=& 
 \Omega\sub{c}^{1/3} \nu\sub{c} 
 \exp\left(-{E\sub{c} \over k T} \right)
 \left[1- \exp\left(-{q\sub{l,c} \Delta T\sub{c}
  \over kT^2} \right) \right],
 \label{growth-c}
\end{eqnarray}
where $E_i$ ($i = {\rm silicate\ (si), \ carbon\ (c)}$) is activation
energy of crystallization, $\Omega_{i}$ is volume of the unit cell,
$\nu_i \sim 10^{13}$\,s$^{-1}$ is vibration frequency of the lattice.
In the square bracket, $q\sub{l,si}$ is latent
heat of crystallization from the melt per unit cell of silicate,
$q\sub{l,c}$ is that of graphitization from amorphous carbon, $\Delta
T_i \equiv T_{{\rm e}, i} - T$ is the supercooling, in other words,
the temperature difference between a temperature $T_{{\rm e}, i}$ of
the transition from the melt or the amorphous state to the crystal in
equilibrium and the temperature $T$ in concern.  The the
square-bracket factors indicate the correction near the melting
temperature and almost equal unity at temperatures except near the
melting temperature.
The heating rates due to deposition of the latent heat of
crystallization are expressed by
\begin{eqnarray}
 \Gamma\sub{si}= - 4 \pi a\sub{si}^2 
 {q\sub{l,si} \over \Omega\sub{si} }{da\sub{si} \over dt},\ \ \
 \Gamma\sub{c}= 4 \pi a\sub{c}^2 
 {q\sub{l,c}  \over \Omega\sub{c} }{da\sub{c} \over dt }. 
\end{eqnarray}

\subsection{ Determination of activation energy of graphitization}

\begin{figure}
\epsscale{.80}
\plotone{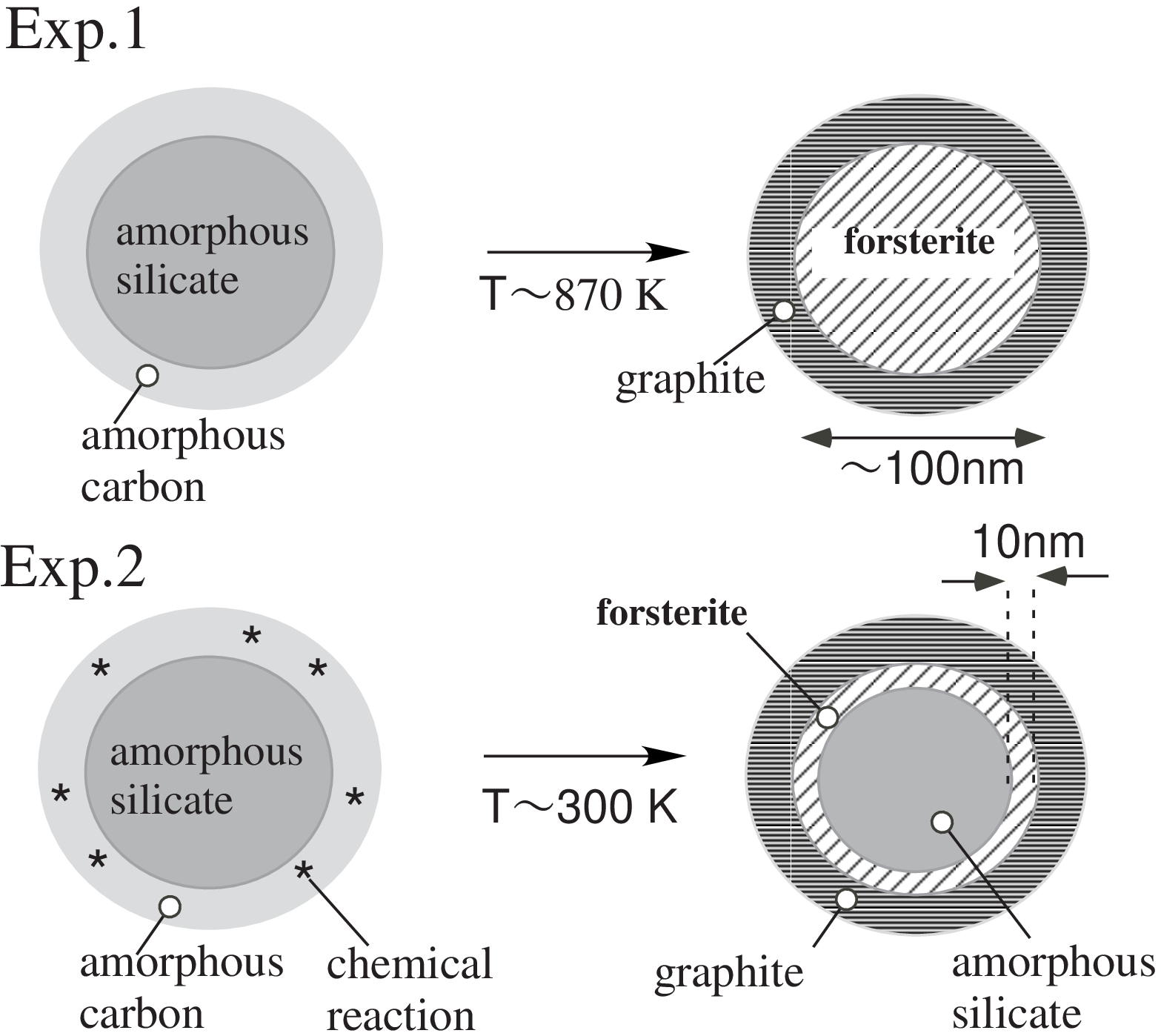}
\caption{Schematic illustration of the crystallization experiments of
 particles having amorphous silicate core and amorphous carbon
 mantle. The particles used in Exp.~1 are those consisting of an
 amorphous silicate of forsterite composition and an amorphous carbon
 layer. Particles used in Exp.~2 are the same as those in Exp.~1 but
 contain reactive molecules (methane in the experiment by
 \cite{kaito2007a}) in the mantle.  
\label{fig1}}
\end{figure}

First we shall derive the activation energies of crystallization in
amorphous silicate and amorphous carbon, (i.e., $E\sub{si}$ and
$E\sub{c}$), by applying the model described in the previous section
to crystallization experiments on submicrometer-sized grains having an
amorphous silicate core and an amorphous carbon mantle. The values of
the activation energies will be used in the analysis of the more
sophisticated experiment of crystallization by \citet{kaito2007a}.
The upper panel of Fig.~\ref{fig1} shows a schematic picture of the
crystallization experiments of amorphous silicate by \citet{kaito2006,
  kaito2007a}. \citet{kaito2006} performed an annealing experiment on
amorphous silicate particles ($\sim$ 50-100\,nm in diameter) covered
with an amorphous carbon layer ($\sim$ 20-30\,nm).  In this experiment
(which we call Exp.~1 hereafter), the crystallization of the amorphous
silicate core of forsterite composition was observed at the ambient
temperature of 870\,K in vacuum, which is 200\,K lower than the case
for bare silicate grains \citep{fabian2000, kamitsuji2005}.  The
amorphous carbonaceous layer was also graphitized.  \citet{kaito2006}
also found that the crystallization in both the silicate core and the
carbon mantle proceeded from the interface between the core and the
mantle.   It might seem curious that the ``crystallization
  temperature'' of 870\,K is lower than the glass transition
  temperature $\sim$ 990 K for amorphous silicate of forsterite
  composition \citep{speck2008}.  Actually, crystallization of the
  amorphous silicate core occurs above the glass transition
  temperature in Exp.~1 as well.  Indeed, our analysis of Exp.~1 shows
  that crystallization of amorphous silicate occurs around 2000\,K in
  a short time because of a rapid temperature elevation in the grain
  due to graphitization of the amorphous carbon mantle followed by the
  cooling (see Fig.~\ref{fig3}).

 Sublimation of the grains is negligible even though they suffer
 temperatures as high as 2000\,K.  We shall show this by comparing
 sublimation and cooling timescales.  The
  sublimation timescale is estimated  by 
 \begin{eqnarray}
 \tau\sub{sub} = \left[ \left( {1 \over a+h} \right)
        \left( d(a+h) \over dt\right) \right]^{-1}
  = {\rho\sub{d} (a+h) \over F\sub{e}},
\label{sublimation}
\end{eqnarray} 
 where $F\sub{e}$ is the sublimation rate given by  
 \begin{eqnarray}
     F\sub{e} =  \alpha P\sub{sat} (T) \sqrt{m \over 2 \pi kT}  
\end{eqnarray}
with $\alpha$ being the evaporation coefficient.  The vapor pressure
$P\sub{sat}$ is approximated by the Clapeyron-Clausius relation expressed 
by $ P\sub{sat} = P_{0} \exp \{ -H /(kT) \}$, where $P_{0}$ is a
constant, $H$ is latent heat of sublimation, and $m$ is mass of
the subliming molecules.   On the other hand, the cooling
timescale including both collisional and radiative coolings is given
by
\begin{eqnarray}
 \tau\sub{cool}
 = {4 \pi (a+h)^3 \rho\sub{d} c\sub{p} T \over 3 (\Lambda\sub{coll}
 + \Lambda\sub{rad})}
  \simeq \frac{2 \rho\sub{d} c\sub{p} (a + h)}
  {3 (n k \bar v+ 2 \sigma\sub{B} \varepsilon\sub{e} T\sub{c}^3) }
\label{tcool2}
\end{eqnarray}  
for $T \gg T_0$. 

\begin{figure}
\epsscale{.80}
\plotone{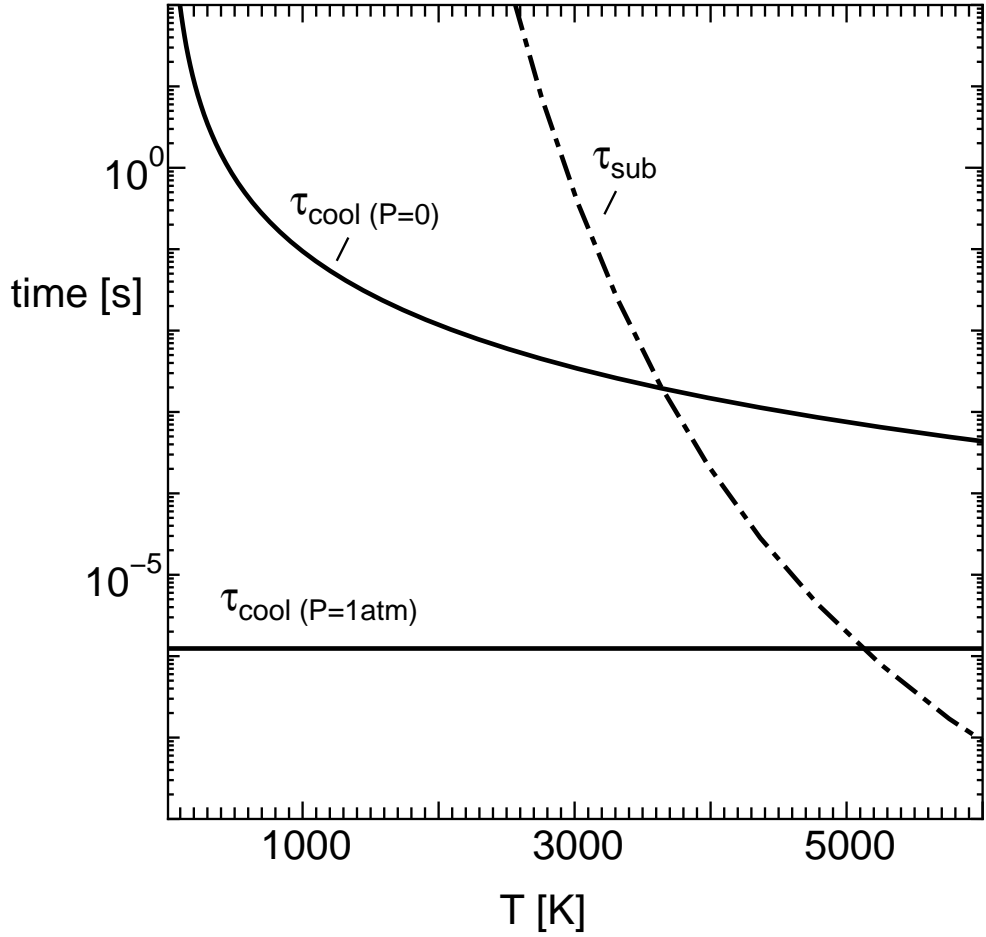}
\caption{Comparison of sublimation timescale $\tau\sub{sub}$
  (dotted-dashed line) and cooling timescale $\tau\sub{cool}$ (solid line)
  of a grain placed at the ambient gas pressure of 1 atm and in
  vacuum.
\label{fig11}}
\end{figure}

\begin{figure}
\epsscale{.80}
\plotone{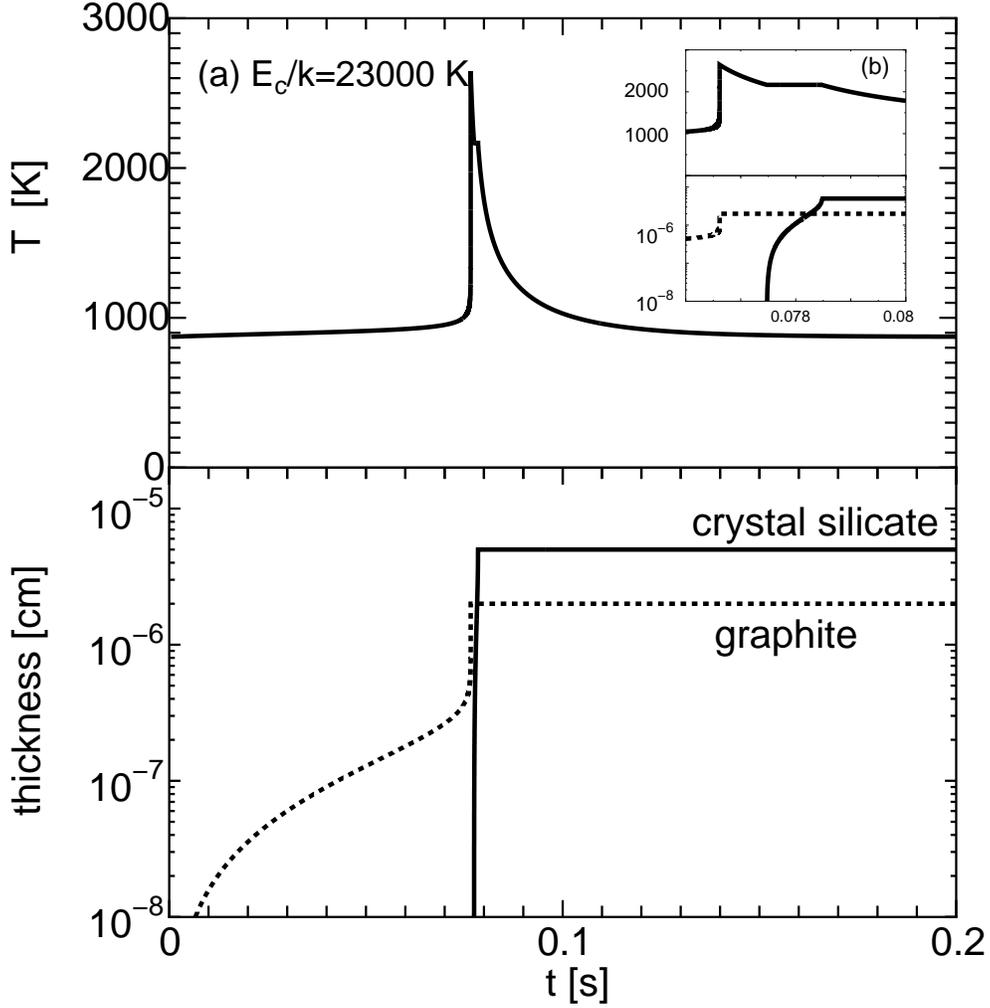}
\caption{Time variations of temperature (upper panel) and thicknesses of
 silicate crystal $l\sub{si}$ and graphite $l\sub{c} $ (lower panel) for
 the setup of Exp.~1. The ambient temperature is $T_0=870\,$K and the
 activation energy of graphitization is taken to be
 $E\sub{c}/k=23000\,$K. The small panel (b) is an enlargement of the
 part of the time interval of $0.076 < t < 0.08\,$s.  \label{fig3}}
\end{figure}

Figure \ref{fig11} compares the sublimation and cooling timescales of
a grain coated with a carbon mantle; the cooling timescales
are estimated for the ambient gas pressure of 1 atm and vacuum.  For
the setup of the Exp.~1 \citep{kaito2006}, we put $a=50\,$nm,
$h=20\,$nm, and $\mu =29$ \citep{satoh2002}. Because of the
core-mantle structure of the grain, sublimation of the carbon mantle
would occur first, for which $H/k=9.5 \times 10^4$ K, $P_{0}=1.3
\times 10^{15}$ dyn cm$^{-2}$ \citep{lide1996}, and $\alpha=1$. We
calculated the emissivity to be $\epsilon = 0.01$ for a core-mantle
grain of $a = 50\,$nm and $h = 20\,$nm using Mie theory.  Figure
\ref{fig11} indicates that the sublimation timescale $\tau\sub{sub}$
is longer than the cooling timescale $\tau\sub{cool}$ for $T \le
3700$\,K in vacuum and $T \le 5200$\,K at 1 atm. The temperature of
the grain experienced during the heating stage is substantially lower
than these temperatures (see Fig.~\ref{fig3} for example),  thus sublimation
is negligible.  This is consistent with the results of the experiments 
 that  the grains suffer little sublimation.

 In this section, we apply the model to Exp.~1 to derive the activation
energy of amorphous carbon.  We solve the energy equation of the
particle given by Eq.~(\ref{exp3-eq}) without chemical reactions in the
carbon mantle ($ \dot\varepsilon=0$) and cooling due to collisions of
surrounding molecules since Exp.~1 was performed in vacuum
($\Lambda\sub{coll} = 0 $).

 There are two cases of crystallization of the silicate core in the
present mechanism, namely, crystallization from amorphous solid or the
melt (liquid).  The latter case occurs if the temperature of the
particle exceeds the melting point by the large heat deposition as will
be shown later. In this case, the melt crystallizes in a subsequent
cooling stage when the temperature becomes lower than the melting
point.   In general, the activation energy for crystallization
$E\sub{si}$ has different values for crystallization from amorphous
solid and liquid. Thus we set
\begin{eqnarray}
E\sub{si} &=& E\sub{si,s} \ \ \mbox{ for  } \ t \le t\sub{m}, \\
E\sub{si} &=& E\sub{si,l} \ \ \mbox{  for  } \ t > t\sub{m},   
\end{eqnarray}
where $t_{\rm m}$ is time when the particle temperature reaches the
melting point, $E\sub{si,s}$ and $E\sub{si,l}$ are the activation
energies for the amorphous silicate and that of the melt, respectively.
We take $E\sub{si,s}/k = 39000\,$K, which was obtained from the analysis
of crystallization experiments of bare amorphous silicate
\citep{fabian2000, kamitsuji2005} (see
also Appendix~\ref{cryst-activation-energy}).  On the other hand, the value
of the activation energy of liquid $E\sub{si,l}/k $ is taken to be
23000\,K \citep{tanaka2008}.  For other parameters, we adopt
$q\sub{l,si}/k = 13800\,$K \citep{navrotsky1995}, $q\sub{l,c}/k
=5100\,$K \citep{kaito2007b}, $T\sub{e,si}= 2160\,$K
\citep{navrotsky1995}, $T\sub{e,c}=3915\,$K \citep{lide1996},
$\Omega\sub{si} = 7.3 \times 10^{-23}\,$cm$^{-3}$ \citep{navrotsky1995},
$\Omega_{\rm c} =8.0 \times 10^{-24}\,$cm$^{-3}$ \citep{lide1996}, and
$\nu\sub{c} = \nu\sub{si} = 1.2 \times 10^{13}\,$s$^{-1}$
\citep{fabian2000}.

Figure~\ref{fig3} 
shows a typical feature of temporal variations in the temperature
$T$ of the particle and the thicknesses $l\sub{si} = a-a\sub{si}$ of the
crystallized layer in the silicate core, and that of the graphitized
layer, $l\sub{c} = a\sub{c}-a$, in the carbon mantle at $T_0=870\,$K. 
Here we take $E\sub{c}/k=23000\,$K, because this value reproduces the
experimental results as will be shown later.

The behavior of the crystallization process is described as follows.
First, graphitization proceeds in the carbon mantle, starting from the
interface between the mantle and the silicate core.  During the growth
of the graphite layer, the latent heat deposited is gradually
accumulated in the particle because the timescale of radiative cooling
($ \simg 1\,$s) is longer than the growth timescale of the graphite
layer ($\simeq 0.1$ to 1\,s) at $T<1000\,$K.  This leads to a gradual
increase in the temperature of the particle.  The temperature rises by
about 100 K from 870\,K to 1000\,K until the thickness $l\sub{c}$ of the
graphite layer becomes a few nano-meters.  
A slight increase in the
temperature at $t = 0.076\,$s leads to rapid graphitization.
 For example, the
growth rate of graphite at 1000\,K is 20 times larger than that at 870 K
( i.e.,~$5 \times 10^{-5}\,$cm\,s$^{-1}$).  This results in a rapid
deposit of the latent heat of graphitization and in turn the 
temperature rise higher than the melting point of forsterite.

The rapid rise in the temperature is a sort of a positive feedback
process, in which the graphitization in the carbon mantle releases the
latent heat deposition and this heating accelerates graphitization.
Once the graphitization of the mantle completes, the particle cools by
thermal emission.  When the temperature decreases below the melting
point of silicate, crystallization of the silicate core begins and
proceeds inward starting from the interface between the silicate core
and the carbon mantle.  The silicate crystallization stops when the
crystallization of amorphous silicate reaches the center of the silicate
core. The timescale of silicate crystallization is about
$10^{-3}\,$s. After the complete crystallization of the amorphous
silicate core, the particle cools further by thermal radiation.

\begin{figure}
\epsscale{.90}
\plotone{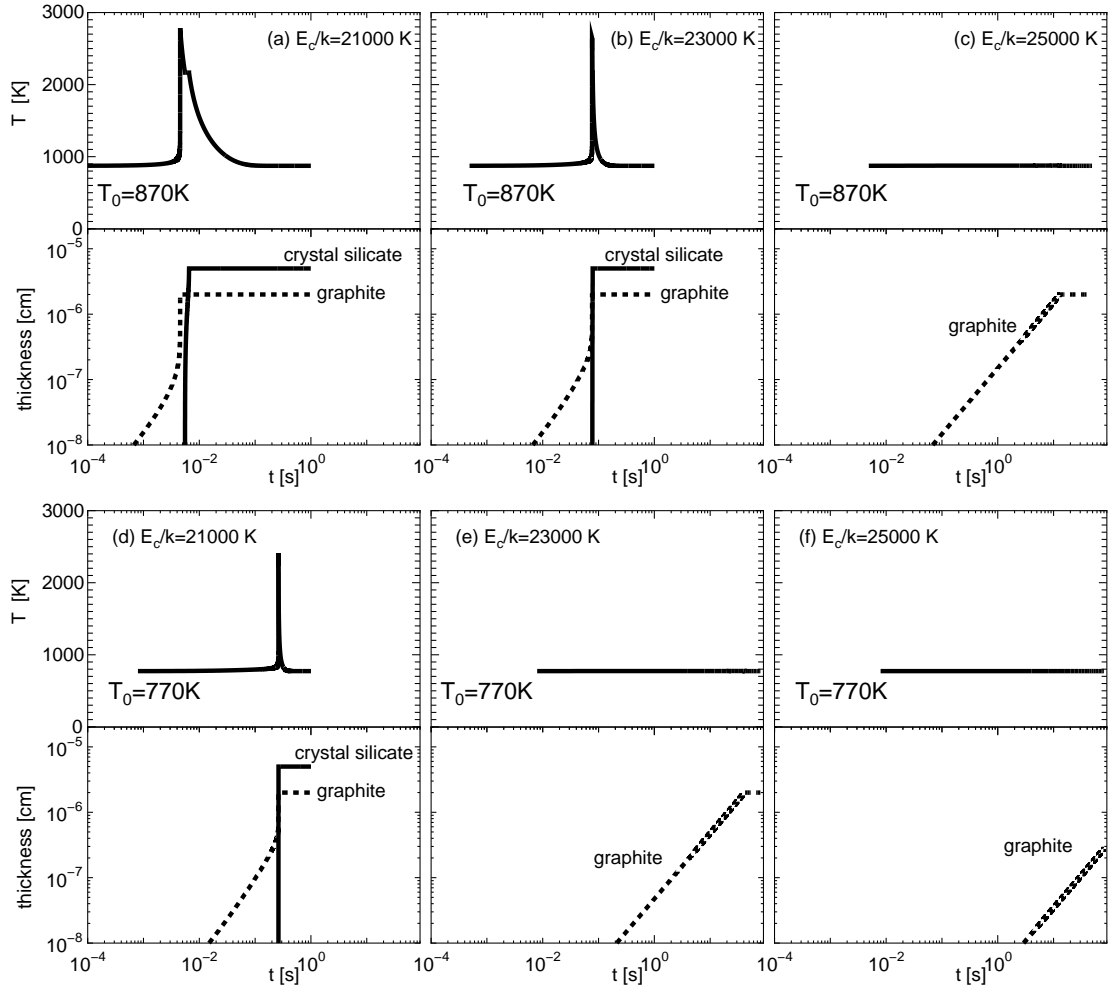}
\caption{Time variations of the temperature and the thicknesses of
silicate crystal and graphite for the setup of Exp.~1 conditions with
varying values of the activation energy of graphitization, $E\sub{c}$,
and the ambient temperature $T_0$.
\label{fig4}}
\end{figure}

Figure~\ref{fig4} shows crystallization features with varying the
values of the activation energy $E\sub{c}$ of graphitization at two
ambient radiation temperatures of $T_0=770$K and 870K.  The result of
\citet{kaito2006} indicates that bare amorphous silicate does not
crystallize at $T_0=770\,$K but does at $T_0=870\,$K. This fact
constrains the value of $E\sub{c}$.  Figures~\ref{fig4}(a)-(c) show
that the $E\sub{c}$-value of $E\sub{c}/k<25000\,$K explains
crystallization of amorphous silicate at $T_0=870\,$K, while
Figs.~\ref{fig4}(d)-(e) show that $E\sub{c}/k>21000\,$K yields no
crystallization at $T_0=770\,$K.  Thus, the activation energy of
amorphous carbon must lie in the range of $21000\, {\rm K} <
E\sub{c}/k < 25000\,{\rm K}$.

The range of the $E\sub{c}$-value is also estimated by comparing the
growth timescale of graphite with the cooling timescale
$\tau\sub{cool}$.  If the growth time is shorter than the cooling time
$\tau\sub{cool}$, amorphous carbon can be graphitized.  Therefore, the
condition of the graphite growth is:
 \begin{eqnarray}
 {h \over da\sub{c}/dt} < \tau\sub{cool},
\label{condi-c}  
 \end{eqnarray}
which leads with the use of Eq.~(\ref{growth-c}) to 
 \begin{eqnarray}
 {E\sub{c} \over k}
 < T \ln \left( \nu\sub{c} \tau\sub{cool} \Omega_{c}^{1/3}
  \over h \right) = 23000 \mbox{K}, 
\end{eqnarray} 
where we set $T = 870\,$K and $\tau\sub{cool} = 1\,$s in obtaining the
value of $E\sub{c}$ on the right-hand side. In the same way, the
condition of no graphite formation at 770\,K is given by $E\sub{c}/k >
19000 \, \mbox{K}$.  The range of $E\sub{c}$ thus estimated is
consistent with the range of $E\sub{c}$ constrained from Fig.~\ref{fig4}.

From detailed calculations as shown in Fig.~\ref{fig4},
 we find that the range of
$E\sub{c}$ to explain the experimental results is 
\begin{equation}
 21300\,{\rm K} < E\sub{c} < 23600\, {\rm K}.
\end{equation}
In what follows, we use $E\sub{c} = 23000 \,{\rm K}$ as a standard
value of the activation energy of graphitization. \\

\section{Low-Temperature Crystallization in Laboratory Conditions}

\subsection{Example of low-temperature crystallization}
\label{LTC-example}
 
\citet{kaito2007a} performed another experiment on crystallization of
amorphous silicate particles coated  with amorphous carbonaceous layers
(see the lower panel of Fig.~\ref{fig1}). In contrast with Exp.~1, however, the
carbon coating was done in a CH$_4$ atmosphere of its pressure of
$10^{-3}\,$Torr (which we call Exp.~2 hereafter).  The sample was
observed by TEM after exposure to the air of the pressure of 1\,atm.
They found crystallization  at room temperature ($\sim
300\,$K) near the interface of the silicate core and the carbon layer.
The crystallization time is less than one minute in the air (Kaito,
private communication). They consider that heat of oxidation of
methane reacting with oxygen in the air graphitized the amorphous
carbon layer and the latent heat of graphitization in turn induces
crystallization of the amorphous silicate core of the particles.

\begin{figure}
\epsscale{.80}
\plotone{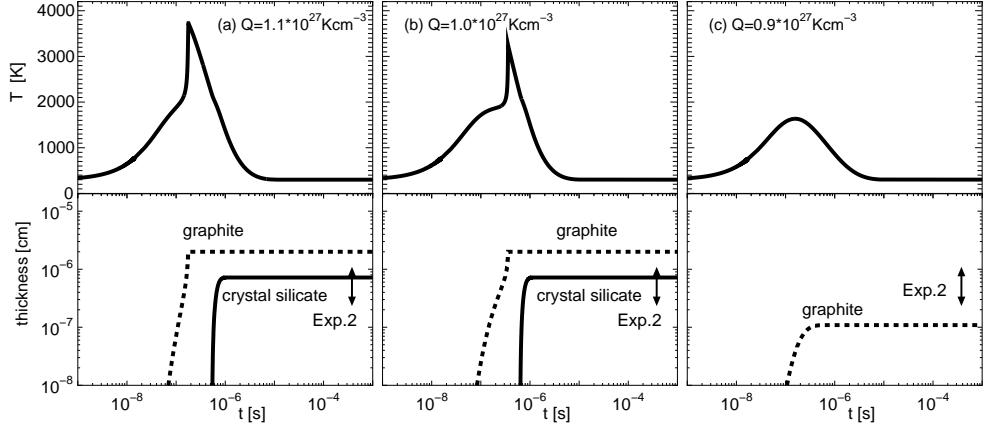}
\caption{Time variations of the temperature and the thicknesses of
 silicate crystal $l\sub{si}$ and graphite $l\sub{c} $ for the setup of
 Exp.~2.  In this figure we set $\tau=5 \times 10^{-8}\,{\rm s}$ and
 varied the $Q$-values: (a) $Q=1.1 \times 10^{27}\,{\rm K\, cm^{{-3}}}$,
 (b) $1.0 \times 10^{27}\,{\rm K\, cm^{{-3}}}$, and (c) $0.9 \times
 10^{27}\,{\rm K\, cm^{{-3}}}$.  The bars with arrows indicate the range
 of the thickness of the crystalline silicate layer observed in Exp.~2
 (Kaito et al. 2007).  \label{fig5}}
\end{figure}

\begin{figure}
\epsscale{.80}
\plotone{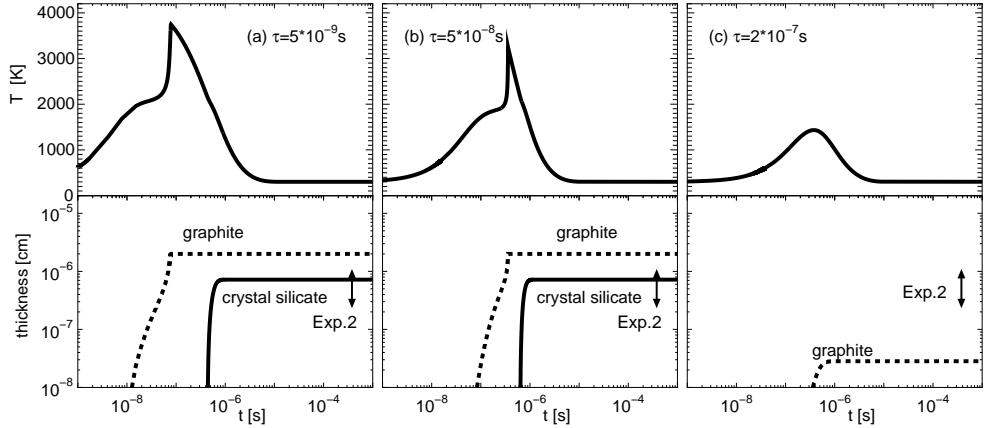}
\caption{The same as in Fig.~5 but for $Q=1.0 \times 10^{27}\,{\rm
 K\,cm^{-3}}$ and (a)~$\tau=5 \times 10^{-9}\,$s, (b)~$5 \times
 10^{-8}\,$s, and (c)~$2 \times 10^{-7}\,$s.  \label{fig6}}
\end{figure}

To clarify the conditions of crystallization due to exothermic
reactions, we solve the equations of the model described in section 2
taking into account exothermic reactions and cooling due to collision
of ambient air molecules at $T_0 = 300\,$K and the pressure of 1 atm.
Figures~\ref{fig5} and \ref{fig6} show a few examples, indicating time
variations of the temperatures and the thicknesses of the crystalline
silicate and graphite layers. In these figures, the radius of a
silicate core and the carbon mantle thickness are the same as in
Exp.~1, (i.e.,~$a = 50\,$nm and $h = 20\,$nm).  Figure~\ref{fig5} are
those for $\tau = 5 \times 10^{-8}$\,s and (a)~$Q = 1.1 \times
10^{27}\,{\rm K\,cm^{-3}}$, (b)~$1.0 \times 10^{27}\,{\rm
  K\,cm^{-3}}$, and (c)~$0.9 \times 10^{27}\,{\rm K\,cm^{-3}}$. In
contrast to the case of Exp.~1, heat of reactions of molecules (${\rm
  CH_4}$ in Kaito et al. (2007)) contained in the carbon mantle with
air leads to gradual rise in the temperature at the first stage and
triggers graphitization of the carbon mantle. Heat released in
graphitization leads further temperature rise above the melting point
of silicate as seen in Figs.~\ref{fig5}(a)-(b).

Silicate crystallization is observed in the subsequent cooling below
the melting temperature.  Note that the timescale of these processes
is much shorter than that in Exp.~1 because the cooling of the
particle is determined by the collisions of air molecules in
Exp.~2. For Fig.~\ref{fig5}(a)-(b), the thickness of crystalline layer
of silicate is about 7 nm.  The thickness of the silicate crystalline
layer is consistent with the experimental results, which are shown by
the bars with arrows in the bottom panels in Fig.~\ref{fig5}.  On the
other hand, no silicate crystallization is observed in
Fig.~\ref{fig5}(c). This is because the stored energy density $Q$ is
insufficient for the temperature to rise above the melting
temperature.  From the calculations with varying $Q$ by a finer step,
we find that $Q \ge 1.0 \times 10^{27}\,{\rm K\, cm^{-3}}$ is
necessary for silicate to crystallize.  In Fig.~\ref{fig6}, we set $Q
=1.0 \times 10^{27}\,{\rm K\,cm^{-3}}$ and varied the reaction
timescale: (a)~$\tau=5 \times 10^{-9}\,$s, (b)~$5 \times 10^{-8}\,$s,
and (c)~$2 \times 10^{-7}\,$s.  We find that $\tau < 5 \times
10^{-8}\,$s is necessary for silicate crystallization.  Both
conditions of $Q \ge 1 \times 10^{27}\,{\rm K\,cm^{-3}}$ and $\tau < 5
\times 10^{-8}\,$s are necessary for silicate crystallization.

\subsection{Analysis of Exp.~2}

We give analytical consideration of Exp.~2 and estimate a minimum value
of the stored energy density $Q$ for inducing crystallization of the
silicate core in the conditions of Exp.~2.  Since cooling of the
particle is determined by collisions of air molecules in Exp.~2, the
timescale of the cooling is given by
\begin{eqnarray}
 \tau\sub{cool}
 = {4 \pi (a+h)^3 \rho\sub{d} c\sub{p} T \over 3 \Lambda\sub{coll}}
  \simeq \frac{2 \rho\sub{d} c\sub{p} (a + h)}{3 n k \bar v} 
 \quad (T \gg T_0), 
\label{tcool}
\end{eqnarray}  
which is about $10^{-6}\,$s at $T \simeq 1000\,$K.  This timescale is
much shorter than the timescale of the crystallization of amorphous
silicate, which is longer than $10^{-5}\,$s (see 
Appendix~\ref{cryst-activation-energy}). Therefore, crystallization from
the solid amorphous silicate is impossible in the conditions of Exp.~2.
However, once amorphous silicate melts, crystallization is possible
since crystallization from the melt is easier than from the amorphous
solid.  The timescale of crystallization from melting particles
$\tau\sub{cry}$ is estimated from
\begin{eqnarray}
 \tau\sub{cry} = {a \over da\sub{si}/dt} 
\end{eqnarray}
of which the minimum value is about $10^{-6}\,$s at $T \sim 2000\,$K.
This timescale is on the same order of magnitude as the cooling time
in Exp.~2.

As shown in the previous section, graphitization of amorphous carbon
expedites crystallization of the silicate core.  The amorphous carbon
mantle crystallizes after the temperature increases due to chemical
reactions in the mantle.  From Eq.~(\ref{exp3-eq}), the time variation of
the temperature before graphitization (i.e.,~$\Gamma\sub{si} =
\Gamma\sub{c} = 0$) is given by
\begin{equation}
 T = T_0
   + \frac{k Q \tau_{\rm cool}}
          {\rho_{\rm d} c_{\rm p} (\tau_{\rm cool} - \tau)}
     \left[1 - \left(\frac{a}{a + h}\right)^3\right]
     (e^{-t/\tau_{\rm cool}} - e^{-t/\tau}),
\label{T(t)}
\end{equation}
where $\tau_{\rm cool}$ is timescale of the collisional cooling given
by Eq.~(\ref{tcool}).  Here we ignore the radiative cooling because the
cooling is mainly due to collisions of air molecules in Exp.~2.  The
temperature given by Eq.~(\ref{T(t)}) increases with time and reaches a
maximum at
\begin{equation}
 t = \frac{\ln (\tau_{\rm cool}/\tau )}
                          {1/\tau - 1/\tau_{\rm cool}}. 
\end{equation}
The maximum temperature $T_{\rm c}$ is given by
\begin{eqnarray}
 T_{\rm c} - T_0  
 &=& \frac{k Q \tau_{\rm cool}}{\rho\sub{d} c_{\rm p} (\tau_{\rm cool} - \tau)}
   \left[1 - \left(\frac{a}{a + h}\right)^3\right]
   \left(\frac{\tau}{\tau_{\rm cool}}\right)^{\tau/(\tau_{\rm cool} - \tau)}
   \left(1 - \frac{\tau}{\tau_{\rm cool}}\right) \nonumber \\
 &\simeq& \frac{k Q}{\rho\sub{d} c_{\rm p}}
   \left[1 - \left(\frac{a}{a + h}\right)^3\right]
 \quad (\tau \ll \tau_{\rm cool}).
\label{tmax}
\end{eqnarray}
%
%
Note that $T_{\rm c}$ is independent of $\tau$ and $\tau_{\rm cool}$ and
determined only by $Q$. Equation~(\ref{tmax}) gives the minimum stored
energy density $Q_{\rm min}$ given a temperature $T_{\rm c}$ that
induces crystallization. Namely, $Q\sub{min}$ is given by
\begin{eqnarray}
 Q\sub{min} \simeq
 { \rho\sub{d} c\sub{p} (a+h)^3 (T\sub{c}-T_0) \over k [(a+h)^3-a^3]}. 
\label{condi-q}
\end{eqnarray}
We get $Q_{\rm min} = 0.8 \times 10^{27}\,$K\,cm$^{-3}$ for
$\tau\sub{cool}=10^{-6}\,$s and $T_{\rm c} = 2000\,$K, which is the
graphitization temperature estimated from Eq.~(\ref{condi-c}), because
graphitization of the carbon mantle leads to silicate crystallization
in Exp.~2. Note that this $Q_{\rm min}$-value is in good agreement with
the numerical calculations given in section \ref{LTC-example}.

\section{Conditions of Low Temperature Crystallization in Astrophysical
 Environments}

\begin{figure}
\epsscale{.55}
\plotone{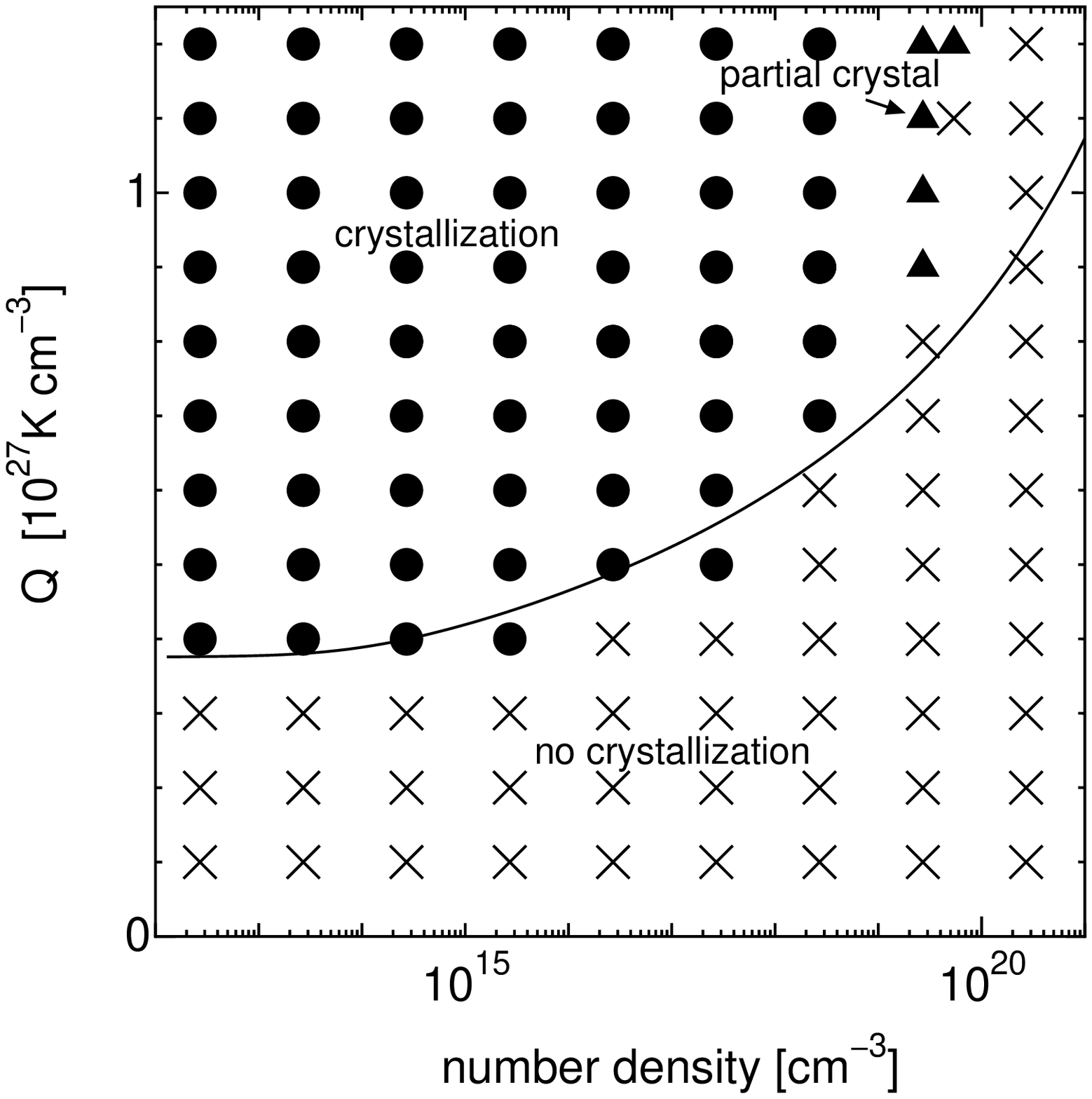}
\plotone{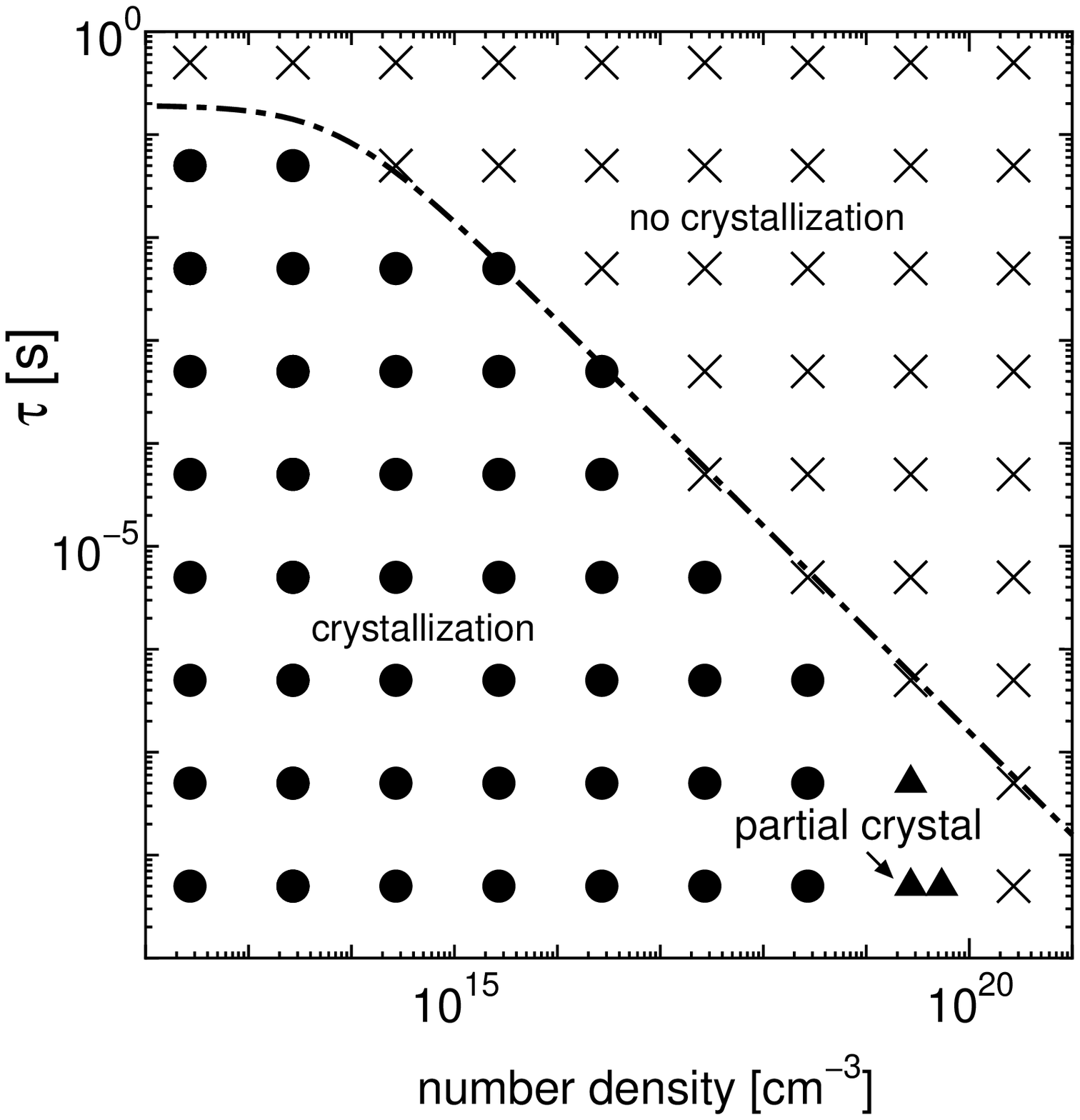}
\caption{Results of calculations  for nonthermal crystallization. 
 Filled circles show  complete crystallizations, 
 triangles show  partial crystallizations,
 and marks $\times$ show no crystallization  of amorphous silicate.
\label{fig7}}
\end{figure}

\begin{figure}
\epsscale{.55}
\plotone{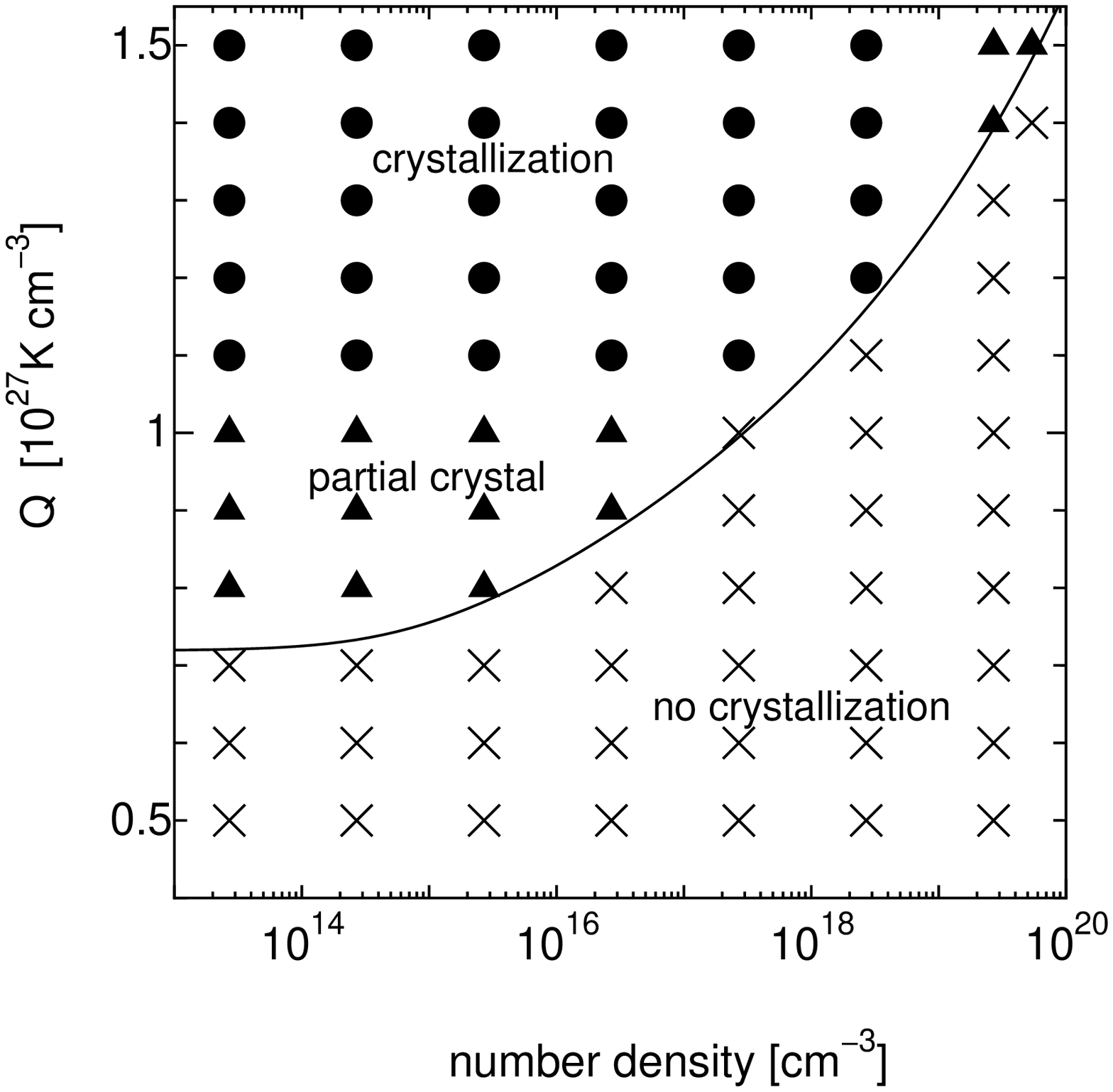}
\plotone{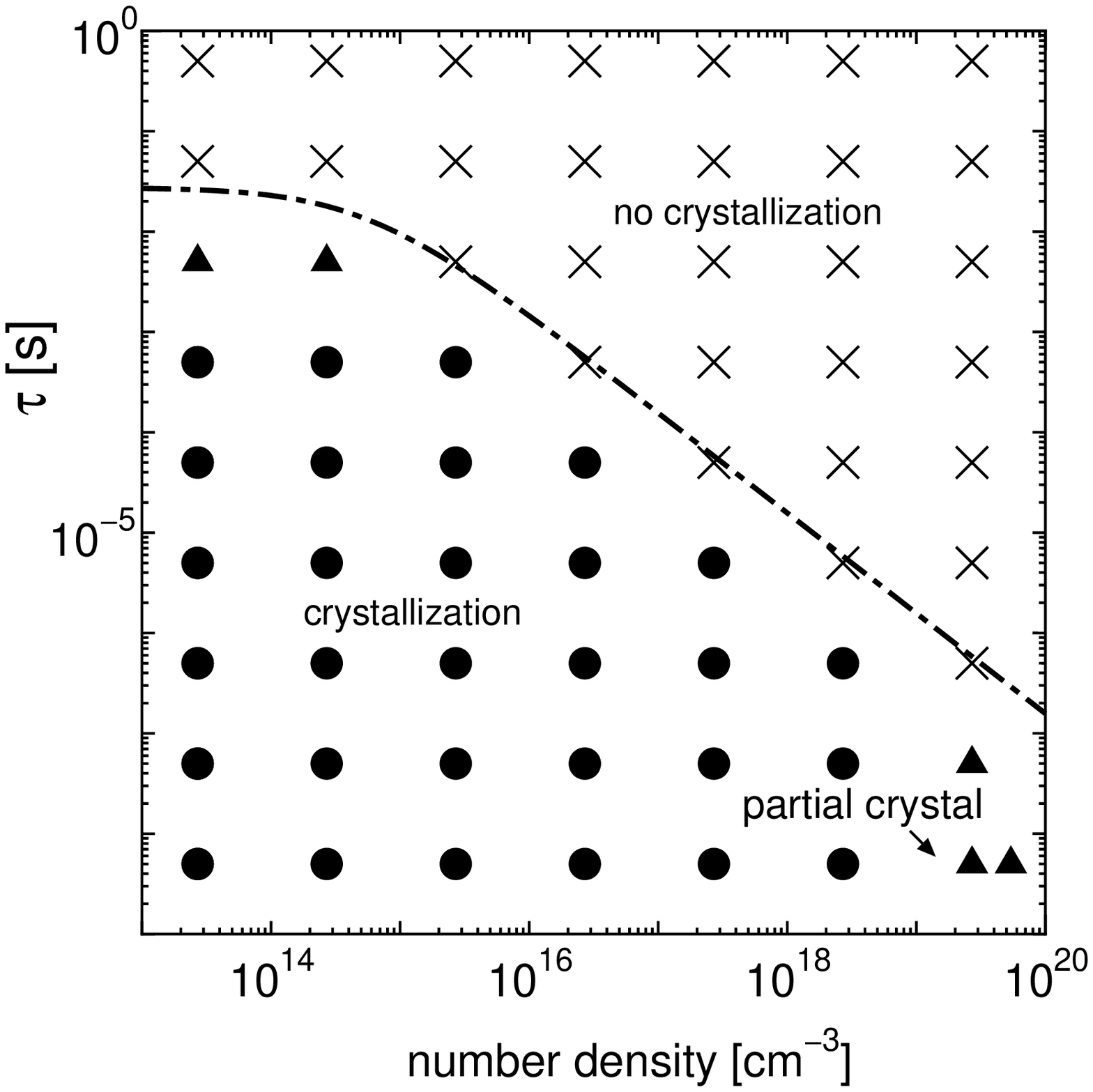}
\caption{
The same as Fig.~\ref{fig7} but for  no graphitization of  amorphous carbon.  
\label{fig8}}
\end{figure}

The previous section focused on the analysis of Exp.~2, in which partial
crystallization of a silicate core was observed when the particle was
exposed to the ambient gas of its pressure of 1\,atm (i.e., $n \sim
10^{19}$\,cm$^{-3}$).  For low gas densities as those in astrophysical
environments, however, crystallization becomes easier than in the
conditions in Exp.~2 because of inefficient collisional cooling by the
ambient gas.
Radiative cooling dominates the collisional cooling for very low gas
densities.  The gas density for which radiative and collisional coolings
become comparable is estimated from $\Lambda_{\rm rad} = \Lambda_{\rm
coll}$ to be
\begin{equation}
 n = \frac{\sigma_{\rm B} \varepsilon_{\rm e}}{k \bar v} T^3
 = 1.4 \times 10^{13}
   \left(\frac{\varepsilon_{\rm e}}{0.01}\right)
   \left(\frac{\mu}{2.2}\right)^{1/2}
   \left(\frac{300 \,{\rm K}}{T_0}\right)^{1/2}
   \left(\frac{T}{1000\,{\rm K}}\right)^3
   {\rm cm^{-3}}
\end{equation}
for $T \gg T_0$, where $\mu = 2.2$ is the mean molecular weight of a gas
of the solar composition \citep{asplund2006}. As will be shown later
(Figs.~\ref{fig7} and \ref{fig8}), complete crystallization of the
silicate core is realized for the gas densities lower than that
estimated above, unless the stored energy density $Q$ is so small or the
reaction timescale $\tau$ is so long to realize the crystallization.

We consider two possible processes of crystallization induced (A)~by
graphitization followed by chemical reactions in the mantle, and
(B)~reactions alone.  The case (B) is considered to clarify the effect
of the reactions without pre-heating played by the amorphous carbon
mantle in Exp.~2. Actual situations may be situated between these two
possibilities depending on the 
 amorphousness and the composition of the
mantle.

First we consider the case (A).  We carry out numerical calculations
with varying the ambient gas density, the stored energy density $Q$, and
the reaction timescale $\tau$ and explore the crystallization
conditions. Figure~\ref{fig7} summarizes the result as a function of the number
density of ambient gas molecules, showing the ranges of $Q$ (the upper
panel) and $\tau$ (the lower panel), in which amorphous silicate
particles crystallize.  It should be noted that, as the gas density
decreases, the regions of $Q$ and $\tau$ inducing the crystallization
widely extend to small $Q$ and long $\tau$. This extension is caused by the
slow cooling due to inefficient collision frequency of the ambient gas.
Complete crystallization of the silicate core is possible for the gas
density $n \le 10^{18}\,{\rm cm^{-3}}$ (depending on the stored energy
density $Q$ and the timescale of the reactions $\tau$); the gas density
in Exp.~2 is found to be marginal and results in partial crystallization
of the silicate core.

Next we explore the case (B) of crystallization induced by chemical
reactions alone.
Figure~\ref{fig8} shows the result of the numerical calculations. The overall
tendency is the same as in the case (A). Namely, the ranges of $Q$ and
$\tau$ yielding the crystallization extend to the regions of small $Q$
and long $\tau$ with decreasing the gas density. Compared to the case
(A) shown in Fig.~\ref{fig7} quantitatively, the minimum value of $Q$ required
for the crystallization increases about twice of that in the case
(A). In addition, partial crystallization occurs even for low gas
densities.  For $n \siml 10^{13}\,$cm$^{-3}$, partial crystallization
occurs in the range of $0.8 \siml Q \siml 1.1 \times 10^{27}\,$K\,cm$^{-3}$,
 while complete crystallization occurs for $ Q \simg 1.1
\times 10^{27}\,$K\,cm$^{-3}$.

We formulate the crystallization conditions that reproduce the
numerical results shown in Figs.~\ref{fig7} and \ref{fig8}.  The
condition for the crystallization to occur is given by
\begin{equation}
 Q > Q_{\rm min} \quad {\rm and} \quad \tau < \tau_{\rm cool},
\label{Q,tau-conditions}
\end{equation}
where $\tau_{\rm cool}$ is the cooling timescale of the grain and $Q_{\rm
min}$ is the minimum stored energy density to induce crystallization.  
%
%
 The minimum stored energy density $Q_{\rm min}$ is given
by Eq.~(\ref{condi-q}), in which $T_{\rm c}$ depends on the case~(A) or (B).
In the case (A), crystallization of the amorphous silicate core is
induced by  graphitization of the amorphous carbon mantle, and the
temperature $T\sub{c}$ is estimated from the graphitization condition
(\ref{condi-c}) expressed as
\begin{eqnarray}
 T\sub{c} = 
 { E\sub{c} \over k }  
 \left[%
  \ln \left\{ {2 \Omega\sub{c}^{1/3} \nu\sub{c} \rho\sub{d} c\sub{p}
    \over 3 (k  n \bar v + 2 \sigma\sub{B} \varepsilon\sub{e} T\sub{c}^3)}
 \left({a \over h} + 1 \right)
 \right\} \right]^{-1}
 \quad (T\sub{c} \gg T_0).
\label{condi-q2}
\end{eqnarray}
The solid line in the upper panel of Fig.~\ref{fig7} shows $Q\sub{min}$ with
$T_{\rm c}$ determined from Eq.~(\ref{condi-q2}). In the case~(B),
$T_{\rm c}$ is estimated from the relation that 
\begin{eqnarray}
 T\sub{c} = 
 { E\sub{si} \over k }  
 \left[%
  \ln \left\{ {2 \Omega\sub{si}^{1/3} \nu\sub{si} \rho\sub{d} c\sub{p}
         \over 3 (k  n \bar v + 2 \sigma\sub{B} \varepsilon\sub{e} T\sub{c}^3)}
 \left(1 + {h \over a} \right)
 \right\} \right]^{-1}
\label{condi-q3}
\end{eqnarray}
derived from the silicate crystallization condition that
\begin{eqnarray}
 {a \over |da\sub{si}/dt|}  \siml \tau\sub{cool}.
\label{condi-si}  
\end{eqnarray}
The solid line in the upper panel of Fig.~\ref{fig8} shows $Q_{\rm min}$ using
Eq.~(\ref{condi-q3}).  One should note that both $Q\sub{min}$ thus estimated
well reproduces the results of the numerical calculations.  It should be
pointed out that $Q\sub{min}$ does not depend on the absolute size of a
grain but on the ratio between the mantle thickness $h$ and the core
radius $a$.  The value of $Q\sub{min}$ decreases with increasing the
ratio $h/a$. The cosmic abundance of elements puts a constraint on the
ratio $h/a$ to range from 0.3 to 0.5 \citep{kimura03a}.  This implies
that $Q\sub{min}$ does not depend much on $h/a$ in astrophysical
situations. We used $h/a=0.4$ as a standard value in this study.

A maximum value of reaction timescale $\tau$ allowing crystallization
is given by $\tau_{\rm cool}$ (see Eq.~(\ref{Q,tau-conditions})).  The
dot-dashed lines in the lower panels of Figs.~\ref{fig7} and
\ref{fig8} show $\tau_{\rm cool}$ given by Eq.~(\ref{tcool2}) as a
function of the gas density $n$. The cooling time becomes constant for
$n \le 10^{13}$ cm$^{-3}$, for which density the cooling is almost due
to thermal radiation from the particles.  The difference in $\tau_{\rm
  cool}$ is small in cases (A) and (B). One should note that
$\tau_{\rm cool}$ reproduces the boundary of the crystallization
region in the $n$-$\tau$ plane, indicating that the condition
Eq.~(\ref{Q,tau-conditions}) works.

\section{Discussion}

We have constructed a theoretical model for low-temperature
crystallization induced by exothermic chemical reactions in
submicrometer-sized grains having a silicate core and a carbonaceous
mantle. The validity of our model is proved by its application to the
results of laboratory experiments on crystallization of amorphous
silicate grains by Kaito and his colleagues. This enables us to
formulate crystallization conditions of amorphous silicate in
low-temperature environments using the stored-energy density $Q$ and the
reaction timescale $\tau$, irrespective of details of chemical reactions
(see (\ref{Q,tau-conditions})). It should be pointed out that the
conditions given in Eq.~(\ref{Q,tau-conditions}) are valid
regardless of the grain structure, in spite of the fact that the
equations are derived from the assumption of a core-mantle structure.
We conclude that the low-temperature crystallization mechanism discussed
in this paper should work at any dust grains unless the density $Q$ of
energy stored in the grains lies below the critical density $Q_{\rm
min}$.

Our results suggest that crystallization of amorphous silicate takes
place in much lower temperatures than hitherto considered. As stated
in section 1, crystallization of amorphous silicate needs high
temperatures ($ \simg 1000\,$K), while the present mechanism works by
moderate heating, say, of a few hundred kelvin, depending the gas
density of interest. In a protoplanetary disk, the number density of
gas and its temperature at 1 AU is estimated to be $n \sim 10^{14}\,
$cm$^{-3}$ and $T = 300\,$K, respectively \citep{hayashi1981}. One
could derive the conditions for inducing silicate crystallization from
Figs.~\ref{fig7} and \ref{fig8} to be $ Q_{\rm min} = (4-8) \times
10^{26}\,$K\,cm$^{-3}$ and $\tau\sub{cool} = 10^{-2}\,$s. This $Q_{\rm
  min}$-value corresponds to the concentration of reactive molecules
in the mantle being (5-10)~\% for the heat of chemical reaction
$E\sub{r} \sim 10^{5}\,$K.  This concentration of reactive molecules
is in harmony with (1-10)~\% suggested in a carbonaceous mantle of
interstellar dust grains \citep{greenberg1976}. Therefore, the
nonthermal crystallization of silicate grains might come into effect
in protoplanetary disks if heat of chemical reactions is deposited on
the grains within a short timescale.

 The minimum stored energy $Q\sub{min}$ measures hardness of
  crystallization of silicate of various composition in the nonthermal
  crystallization; the larger $Q\sub{min}$, the harder the
  crystallization is.  When the crystallization is induced by
  reactions (case (B) in section 4), $Q\sub{min}$ is determined mainly
  by the activation energy of crystallization of the silicate (see
  Eq.(\ref{condi-q}) and Eq.(\ref{condi-q3})).  For example, the
  activation energies of crystallization of the amorphous silicate are
  $E\sub{si}=39000$\,K for forsterite composition
  and $E\sub{si}=42000$\,K for that of enstatite composition
  \citep{fabian2000}.  The
  $Q\sub{min}$ values are $ Q_{\rm min} = 8 \times
  10^{26}\,$K\,cm$^{-3}$ for forsterite composition and $ Q_{\rm min}
  = 9 \times 10^{26}\,$K\,cm$^{-3}$ for enstatite composition.  This
  indicates that the crystallization of amorphous silicate of
  enstatite composition is harder than that of forsterite composition.
  On the other hand, in the case that silicate crystallization is
  induced by crystallization of the amorphous mantle (graphitization
  for an amorphous carbon mantle) due to deposition of heat of
  reactions in the mantle (case (A)), $Q\sub{min}$ is determined by
  the activation energy of crystallization of the mantle material.  In
  this case,  the hardness of silicate crystallization is independent of
  its composition.

As long as there is a mechanism to trigger chemical reactions in a  
carbonaceous layer of dust grains, amorphous silicate in the grains  
would suffer from low-temperature crystallization. One of the  
plausible mechanisms is a flash heating of dust by shock waves, which  
has been proposed as a formation mechanism for chondrules found in  
meteorites. Once the dust is heated moderately in a shock region,  
diffusion of reactive molecules becomes active and will lead to chain  
chemical reactions. Note that the heating required for inducing  
crystallization of amorphous silicate is much lower than the one for  
chondrule formation. Collisions among dust particles, which raise the  
temperature in a local region of their surfaces ($\sim$ tens of  
degree), may be another mechanism of flash heating. Since the 
above-mentioned mechanisms are most likely common to happen in  
protoplanetary disks, such a flash heating would trigger silicate  
crystallization in a protoplanetary disk.

Even if shock waves and mutual collisions are not effective enough to  
trigger chemical reactions, the low-temperature crystallization of  
amorphous silicate could be induced by spontaneous heating known as  
the Wigner storage mechanism in nuclear reactor engineering  
\citep{shabalin2003}. It is well known for graphite irradiated by  
neutrons that energy stored in the form of lattice defects will be  
released spontaneously when the number density of the defects reaches  
a certain value. A similar phenomenon of spontaneous stored-energy  
release was observed in experimental studies of neutron-irradiated  
ices \citep{carpenter1987, shabalin2003, kulagin2004}.  
\citet{carpenter1987} found that the temperature of a cold solid  
methane moderator rises rapidly and uncontrollably when the heat loss  
becomes less effective after irradiation of fast neutrons into the  
moderator. The rapid temperature rise is due to the phenomenon that  
the energy stored at low temperatures in the form of reactive  
molecules was abruptly released from the irradiated methane by  
thermally activated diffusion of the molecules and their subsequent  
reactions. 

Reactive molecules in the ice mantle of grains would form by irradiation
of UV or high energy particles such as cosmic ray in the outer cold
region of a protoplanetary disk or in a molecular cloud. When reactive
molecules in such a grain are accumulated to a critical number density,
the heat released by the activation of reactions would induce
crystallization of amorphous silicate in the grain. However, there is no
observational evidence for the presence of crystalline silicate in the
outer cold region of a protoplanetary disk nor in a molecular
cloud. This implies either that there is an observational selection
effect to obscure the presence of crystalline silicate in those
environments or that the temperatures are too low to activate chain
chemical reactions in the outer cold region of a protoplanetary disk and
in a molecular cloud.  If the latter is the case, observations with high
spatial resolutions would constrain the conditions to trigger silicate
crystallization and the accumulation rate of reactive molecules in
low-temperature astrophysical environments.

\acknowledgments
We are grateful to C. Kaito and  Y. Kimura for helpful comments 
 and discussion    
and an anonymous reviewer for useful comments to improve the paper.  
T. Y. acknowledges support by the Grant-in-Aid for Scientific Research
(21244011) from JSPS.  H. K. is supported by the grants from CPS, JSPS,
and MEXT Japan.

\appendix

\section{Determination of the Activation Energy of Crystallization}
\label{cryst-activation-energy}
Crystallization of an amorphous solid occurs through nucleation and
growth of a crystalline nucleus.  Empirically, a characteristic time
scale $\tau$ of crystallization of amorphous silicate placed at
temperature $T$ is defined by
\begin{eqnarray}
\tau= \nu\sub{si}^{-1} \exp \left( E \over kT \right),
\label{empi}
\end{eqnarray}
where $E$ is the effective activation energy
 \citep{hallenbeck1998, fabian2000}. 

There are two kinds of nucleation: one is homogeneous
nucleation and the other is heterogeneous nucleation in which impurities
or contact interfaces act as seed nuclei for crystal growth. Homogeneous
nucleation occurs in the absence of the impurities or the interfaces.
The timescale of crystallization due to homogeneous nucleation is
controlled by the rate of nucleation of crystalline nuclei, $J$, and the
growth rate $w$ of the crystal, and is given \citep{kouchi} by
\begin{eqnarray}
 \tau\sub{homo} = 
   \left(  3 \over \pi J w^3  \right)^{1/4},
\label{time}
\end{eqnarray}
where $J$ is a function of the interfacial energy $\sigma$ between
amorphous solid and crystals, and the activation energy of molecular
diffusion $E\sub{si}$ is expressed \citep{seki1981,
 kouchi,  tanaka2008}  as 
\begin{eqnarray}
  J =  { 4 \pi   \nu\sub{si} \over 3 \Omega\sub{si}}\sqrt { 
  \beta \sigma \over \pi kT }
 \exp\left\{-{E\sub{si} \over kT}
  -{ 4 (\beta \sigma)^3 T  
 \over 27 q\sub{l,si}^2 k(\Delta T\sub{si})^2 }  \right\} .
\end{eqnarray}
Here, $\Delta T\sub{si} \equiv T\sub{e,si} - T$ is the supercooling,
$\beta $ is the geometrical parameter depending on the shape of
crystalline nuclei (4.8 for spherical nuclei).  The growth rate of the
crystal $w$ is given by Eq.~(\ref{growth-si}),  i.e., $w=-da\sub{si}/dt$. 

For heterogeneous nucleation, the timescale of crystallization depends
on the number of the nuclei in the amorphous solid. A picture of
amorphous solid is that it consists of an ensemble of microcrystals,
which have sizes of the lattice scale and are distributed in random
orientations. According this view, there are already enough number of
crystalline nuclei, and the timescale of (macroscopic) crystallization
is given by
\begin{eqnarray}
 \tau\sub{hetero}
 = \frac{\Omega^{1/3}}{w}. 
\end{eqnarray}
In this case, $E\sub{si}$ corresponds to the effective activation energy
$E$ in Eq.~(\ref{empi}) if one ignores the melting ($e^{-q\sub{l,si}
\Delta T\sub{si}/kT^2} \ll 1$).

\begin{figure}
\epsscale{.80}
\plotone{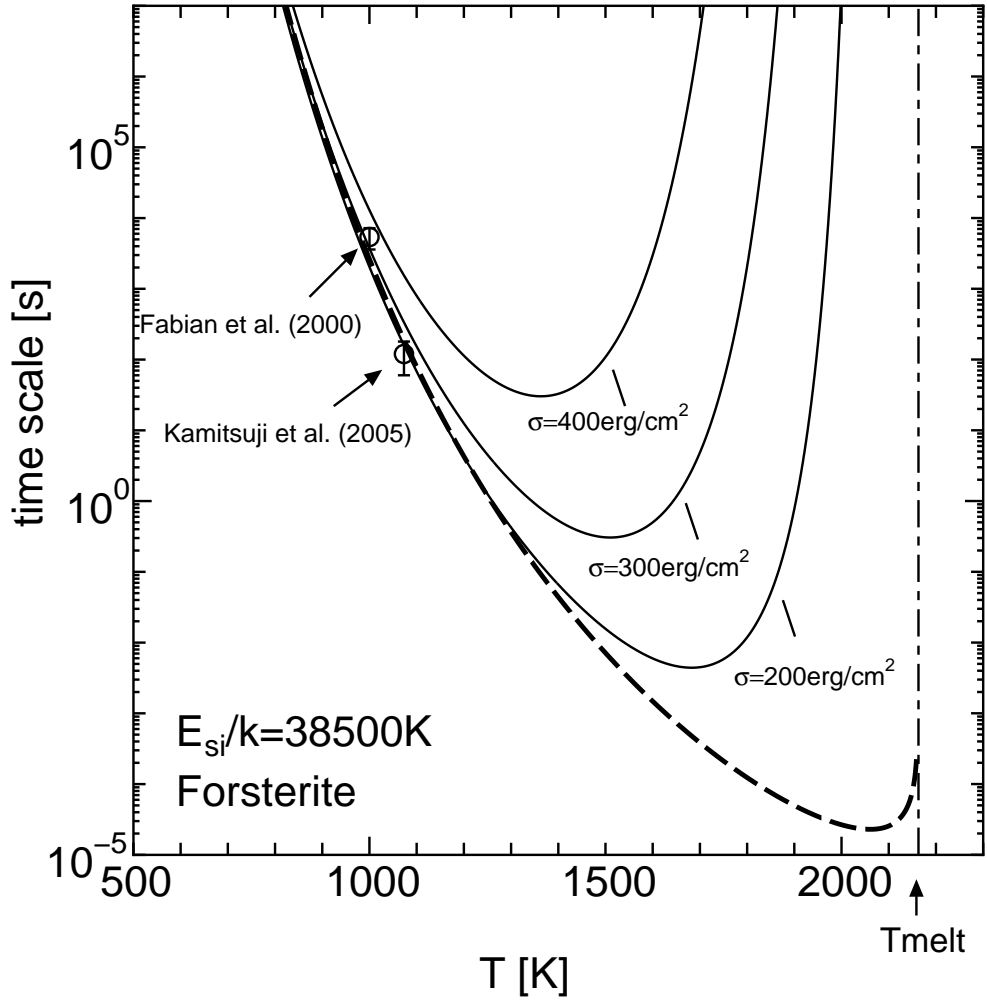}
\caption{Crystallization time as a
function of temperature $T$.  Also plotted are the results of the
experiments by \citet{fabian2000} and \citet{kamitsuji2005}.
\label{fig9}}
\end{figure}

Figure~\ref{fig9} shows the crystallization timescales as a function of
temperature.  From the figure, we put constraints on the values of
$E\sub{si}$ and $\sigma$. Namely, we find that $E\sub{si} /k \simeq
38000- 39000\,$K and $\sigma < 400\,{\rm erg\, cm^{-2}}$ satisfy the
results of the experiments for homogeneous nucleation, while
$E\sub{si}/k \simeq 38500\,$K
satisfies the results of the experiments for heterogeneous nucleation.
Note that the close value of the activation energy is yielded if we
assume homogeneous nucleation or heterogeneous nucleation.  This is
because the timescale of the crystallization is mainly controlled by
crystalline growth rather than nucleation in the relevant
temperatures.  The present value of $E\sub{si}$ is in good agreement
with the effective activation energy ($E\sub{si} /k =39100 \pm 400\,$K)
estimated by \citet{fabian2000}.

\clearpage



\clearpage

\end{document}